\documentclass[12pt,letter,logo,copyright,nonumbering]{xds}

\usepackage{subcaption}
\usepackage[sort]{natbib}
\usepackage{dblfloatfix}
\usepackage{ulem}
\usepackage{float}
\usepackage{caption}
\usepackage{dramatist}
\usepackage{xspace}
\usepackage{pifont} 
\usepackage{multirow}
\usepackage{tcolorbox}
\usepackage{xltabular}
\usepackage{longtable}
\usepackage{mdframed}
\usepackage{hyperref}
\usepackage{tocloft}
\usepackage{amsfonts}
\usepackage{amsmath}
\usepackage{amssymb}
\usepackage{lineno}
\usepackage{multirow}
\usepackage{adjustbox}

\usepackage[bottom]{footmisc}

\usepackage{CJKutf8}
\usepackage{setspace}

\usepackage{graphicx}

\usepackage{dsfont}
\usepackage{array} 
\usepackage{tabularx} 
\usepackage{xcolor} 

\usepackage{lipsum}  
\usepackage{multicol} 

\makeatletter
\def\@BTrule[#1]{%
  \ifx\longtable\undefined
    \let\@BTswitch\@BTnormal
  \else\ifx\hline\LT@hline
    \nobreak
    \let\@BTswitch\@BLTrule
  \else
     \let\@BTswitch\@BTnormal
  \fi\fi
  \global\@thisrulewidth=#1\relax
  \ifnum\@thisruleclass=\tw@\vskip\@aboverulesep\else
  \ifnum\@lastruleclass=\z@\vskip\@aboverulesep\else
  \ifnum\@lastruleclass=\@ne\vskip\doublerulesep\fi\fi\fi
  \@BTswitch}
\makeatother

\addto\extrasenglish{
}

 {\begin{list}{}%
         {\setlength{\leftmargin}{#1}}%
         \item[]%
 }
 {\end{list}}

\reportnumber{001} 
\renewcommand{\today}{}

\usepackage{import}
\usepackage{hyperref}
\usepackage{authblk}
\usepackage{footmisc}

\newcommand{\sys}{\textsf{ExpertWeave}\xspace}

\correspondingauthor{Equal contribution.}

\begin{document}

\title{\vspace{-0.2in}\centering \sys: Efficiently Serving Expert-Specialized Fine-Tuned Adapters at Scale}

\author[1,*]{Ge Shi}
\author[1,*]{Hanieh Sadri}
\author[1]{Qian Wang}
\author[2]{Yu Zhang}
\author[1]{Ying Xiong}
\author[1]{Yong Zhang}
\author[1]{Zhenan Fan}
\affil[1]{Huawei Technologies Canada}
\affil[2]{Huawei Cloud}

\date{}

\begin{abstract}

Expert-Specialized Fine-Tuning (ESFT) adapts Mixture-of-Experts (MoE) large language models to enhance their task-specific performance by selectively tuning the top-activated experts for the task. Serving these fine-tuned models at scale is challenging: deploying merged models in isolation is prohibitively resource-hungry, while existing multi-adapter serving systems with LoRA-style additive updates are incompatible with ESFT's expert-oriented paradigm. We present \sys, a system that serves multiple ESFT adapters concurrently over a single shared MoE base model, drastically reducing the memory footprint and improving resource utilization. To seamlessly integrate into existing inference pipelines for MoE models with non-intrusive modifications and minimal latency overhead, \sys introduces a virtual-memory-assisted expert weight manager that co-locates base-model and adapter experts without incurring memory overhead from fragmentation, and a fused kernel for batched rerouting to enable lightweight redirection of tokens to the appropriate experts at runtime. Our evaluations show that \sys can simultaneously serve multiple adapters of a 16B MoE model on a single accelerator where the baseline runs out of memory, or provides up to $94\times$ more KV cache capacity and achieves up to 18\% higher throughput while using comparable resources, all without compromising model accuracy. \sys maintains low overhead even when scaling to 20 adapters, with a 4--11\% latency increase compared with serving the base model alone.
Source code will be released soon.

\end{abstract}

\maketitle

\newpage
\tableofcontents
\newpage

\section{Introduction}
\label{sec:intro}

Large Language Models (LLMs) have recently gained significant attention for their remarkable performance across a broad range of tasks~\cite{radford2018improving,bai2023qwen,meta2023llama,team2023gemini}. 
These models are typically built using the Transformer architecture with multi-head self-attention and feed-forward network (FFN) layers~\cite{vaswani2017attention}.
Traditionally, FFN layers implement a dense architecture, where a single set of parameters is shared across all tokens to compute their activations. This design simplifies the model architecture and the training process. However, since every token is processed by the same weights, this design lacks specialization across different token contexts and imposes substantial computational cost during training and inference.

Mixture-of-Experts (MoE) architectures~\cite{jacobs1991adaptive,jordan1994hierarchical,shazeer2017outrageously} introduce sparsity into transformer-based LLMs by replacing each FFN layer with a set of \textit{expert}s, where each expert is a small-scale FFN layer. A learned token router activates only a small number of experts for each input token, routing different tokens to different experts in the same layer.
This sparse, conditional computation results in LLMs with significantly more parameters than their dense counterparts~\cite{fedus2022switch,du2022glam,zoph2022designing,lepikhin2020gshard}, while the computation required per token remains bounded by the number of experts activated.

As MoE models gain wider adoption, efficiently adapting them to downstream tasks becomes increasingly important. 
As full-parameter fine-tuning is memory- and compute-intensive and oftentimes prohibitively costly, Parameter-Efficient Fine-Tuning (PEFT) methods are often preferred to save time and cost. One common PEFT method is Low-Rank Adaptation (LoRA)~\cite{hu2022lora}, which introduces trainable low-rank matrices as add-on adapters into self-attention and FFN layers in the model. Although LoRA is effective for dense models, its application in MoE models remains underexplored~\cite{wang2024let}.

To address this gap in MoE models, Wang et al. proposed Expert-Specialized Fine-Tuning (ESFT)~\cite{wang2024let}, a parameter-efficient approach tailored for the MoE architecture. ESFT recognizes an \textit{expert specialization} pattern in MoE models: expert routing concentrates on a small fixed subset of experts for a given downstream task, while the sets of activated experts vary substantially across tasks. Instead of updating all experts, ESFT leverages this pattern to compute per-layer expert relevance on a small sample of task data and select only the top-activated experts with a cumulative relevance score exceeding a threshold $p$ for fine-tuning. Unlike LoRA with a static shape, the number of ESFT experts fine-tuned per layer in one adapter can be different due to different task relevance distributions.

Despite the advantages of ESFT, it introduces practical challenges for inference and serving, especially in modern cloud Model-as-a-Service (MaaS) environments. To deploy fine-tuned models, people often must follow a merge-and-serve approach, in which an adapter is merged into the base model to produce a standalone checkpoint for deployment. This strategy scales poorly in the cloud, as each fine-tuned model must be deployed independently, imposing additional memory and compute overhead.

With dense models, systems like Punica~\cite{chen2024punica} and S-LoRA~\cite{sheng2024slora} have enabled the concurrent inference of multiple LoRA adapters over a shared base model, trading minor latency overhead for drastically reduced deployment cost. However, enabling shared inference for ESFT adapters on MoE models remains a challenge, as ESFT adapters require a fundamentally different inference process than the LoRA-style additive updates.
 
Serving multiple ESFT adapters in one system presents several key challenges. First, integrating multiple adapters, each with its own set of fine-tuned experts, requires the system to incorporate the dynamically loaded adapter experts into the execution pipeline without modifying core computational kernels. Second, memory management for the accelerator becomes much harder as the number of fine-tuned experts varies both across adapters and across layers within an adapter, leading to fragmentation and inefficient utilization of limited device memory. Third, batching requests across multiple adapters that involve both base-model and adapter experts requires precise routing of hidden states so that each token is processed by the correct set of experts. Finally, the routing of tokens must not be a performance bottleneck, particularly during the prefill phases due to the need to efficiently route a large volume of tokens to their corresponding experts.

\begin{figure}
    \centering
    \includegraphics[width=0.476\textwidth]{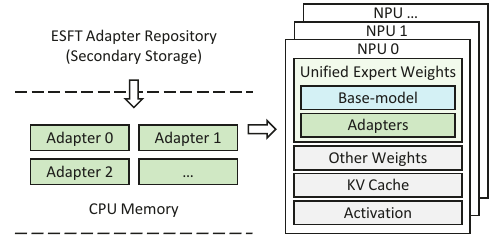}
    \caption{Overview of \sys. ESFT adapters stored in secondary storage are loaded and cached in CPU main memory before loading onto the NPU. \sys manages both base-model and adapter expert weights in a unified manner, ensuring minimal modifications to existing inference pipelines for MoE models.}
    \label{fig:sys_arch}
\end{figure}

To address the aforementioned challenges, we present a system for shared inference of ESFT adapters on MoE models, enabling multiple models fine-tuned on different downstream tasks to be served concurrently and efficiently over a single shared MoE base model.
We make the following contributions:
\begin{itemize}
\item We present \sys, a system capable of serving multiple ESFT adapters concurrently over a shared base model on Ascend NPUs~\cite{liao2021ascend}. An overview of \sys is shown in Figure~\ref{fig:sys_arch}. We demonstrate that, as the number of ESFT adapters increases, our system remains scalable with minimal impact on key performance metrics like throughput and latency.
\item We propose a unified and memory-efficient expert weight management unit that minimizes fragmentation by leveraging virtual memory on Ascend NPUs, sustaining high memory utilization even under heterogeneous adapter expert configurations without requiring modifications to the core computational kernels.
\item We implement an efficient adapter-aware inference runtime with a batched rerouting operator that enables lightweight token dispatching to the correct base-model and adapter experts without hurting latency.
\item We validate the effectiveness of \sys through experiments on end-to-end serving of multiple ESFT adapters, demonstrating scalability, efficiency, effective memory management, and serving accuracy across tasks.
\end{itemize}

\section{Background}
\label{sec:background}

\subsection{Mixture-of-Experts LLMs}

Mixture-of-Experts (MoE) models replace the standard feed-forward network (FFN) layer with an MoE layer that contains a fixed number of experts~\cite{shazeer2017outrageously,jiang2024mixtral}. 
Each expert is a small FFN consisting of linear projections and a nonlinear activation. 
The MoE layer processes flattened inputs of shape $[B, H]$, where $B$ denotes the number of tokens in the batch and $H$ is the hidden dimension.
A learned router $g(x)$ selects the top-$k$ relevant experts for each token $x$ and performs computation only using the activated experts. 
This sparse activation pattern increases model capacity and improves computational efficiency by activating only a small, relevant subset of parameters per token.
DeepSeekMoE~\cite{dai2024deepseekmoe,deepseek_v2,deepseek_v3} further improves model performance by using many fine-grained experts, enabling more flexible expert combinations while maintaining computational efficiency.

Inference engines commonly store expert weights in the MoE layer by stacking each linear layer as a three-dimensional tensor of shape $[M, H_{\text{out}}, H_{\text{in}}]$, where $M$ is the number of experts, and $H_{\text{in}}$ and $H_{\text{out}}$ are the input and output hidden dimensions of a linear layer in an expert~\cite{kwon2023vllm,vllm-ascend}.
The router computes top-$k$ expert IDs of shape $[B, k]$ to determine the selected experts for each token.
The system then dispatches tokens to their selected experts by replicating, permuting and grouping the tokens by the top-$k$ IDs; tokens targeting the same expert are stored in a contiguous chunk.
Batched expert computation is then performed by invoking the Grouped Matrix Multiplication (GMM) operator~\cite{ascend_npu_grouped_matmul_2025} on the permuted tokens and the stacked expert weight tensor. 
The resulting outputs are then combined using the computed weights from the router and unpermuted back to the original order for subsequent layers.

\subsection{Expert-Specialized Fine-Tuning}
LLMs can be adapted to downstream tasks through fine-tuning, where pretrained model parameters are further updated to improve the task-specific performance. In traditional full-parameter fine-tuning, all weight parameters are updated; in contrast, PEFT methods~\cite{han2403parameter} modify only a small subset of parameters while leaving most of the backbone model froze, often by inserting lightweight adapter modules. PEFT methods are usually employed thanks to their improved fine-tuning efficiency with reduced memory and compute requirements.

Expert-Specialized Fine-Tuning (ESFT)~\cite{wang2024let} is a recent PEFT method tailored for the MoE architecture by exploiting the expert specialization pattern in MoE models: for a given task, expert activations are condensed on a small yet largely fixed subset of experts, while the top-activated expert sets differ substantially across different tasks. This pattern motivates ESFT to selectively fine-tune the experts most relevant to the downstream task.
ESFT first samples a subset of task data and uses it to compute a relevance score for each expert.
ESFT proposes two metrics of expert relevance scores: average gate score and token selection ratio.
ESFT then identifies the top-activated experts in each layer by finding the set of experts where the cumulative relevance score exceeds a threshold hyperparameter $p$. These experts are then selected as candidates for fine-tuning on the downstream task. 

During fine-tuning, tokens might still be routed to any expert, but only the selected experts are updated; all the other experts and modules remain frozen. In particular, the router is not updated, opening the possibility for shared inference. In practice, ESFT can achieve performance close to or on par with full-parameter fine-tuning by only fine-tuning roughly 5--15\% of the total experts across all layers, yielding significant savings in resource usage and making it an attractive PEFT method in customizing MoE models.

\subsection{Multi-LoRA Serving}
As LoRA gains adoption as an efficient PEFT technique, the increasing demand for deploying LoRA adapters motivates the emergence of multi-LoRA serving, where a single base model is shared among multiple LoRA adapters. These LoRA adapters are mounted to the system as low-rank matrices alongside the base model. During inference, the system first computes the base-model activation $y=xW$ and then applies the adapter-specific updates $y'= y+xA_iB_i$. Sharing the base model reduces memory pressure when many adapters are active, at the cost of additional computation and latency overhead for the additive adapter updates.

To make multi-LoRA serving efficient in practice, recent systems rely on custom CUDA kernels that batch the LoRA computations across requests and adapters.
Punica~\cite{chen2024punica} introduces a Segmented Gather Matrix-Vector Multiplication (SGMV) kernel that groups tokens corresponding to the same adapter into contiguous segments so a single kernel invocation can process all adapters in parallel. S-LoRA~\cite{sheng2024slora} further enables serving LoRA adapters with varying ranks: its Unified Paging places paged LoRA weights and KV cache in a single memory pool, supporting adapters of different ranks without the need for padding. To enable LoRA computation on paged, non-contiguous adapter weights, S-LoRA proposes a Multi-size Batched Gather Matrix-Matrix (MBGMM) kernel for prefill and a Multi-size Batched Gather Matrix-Vector (MBGMV) kernel for decode. Both systems depend on specialized computational kernels to support multi-LoRA inference efficiently.

Although effective in multi-LoRA setups, these mechanisms cannot be directly transferred to serving ESFT adapters on MoE models. As ESFT fine-tunes selected experts within each MoE layer, the required pattern becomes expert routing followed by expert computations through the GMM operator rather than additive updates from mounted low-rank matrices through SGMV, MBGMM or MBGMV. To the best of our knowledge, there are no analogous multi-adapter serving solutions for ESFT.

\section{Challenges with ESFT Experts}
\label{sec:challenge}

\subsection{Expert Weight Management}
\label{subsec:memory_challenge}

We aim to serve multiple ESFT adapters on a single base MoE model.
The experts to be fine-tuned by ESFT are selected based on the expert relevance scoring mechanism: only the set of top-activated experts with a cumulative relevance score exceeding a threshold hyperparameter $p$ is fine-tuned~\cite{wang2024let}. A direct consequence of this score-based selection is that the number of fine-tuned experts can vary for different layers in one adapter or between different adapters, which complicates expert weight management.

To load the fine-tuned expert weights into NPU memory, a straightforward design is to maintain adapter weight tensors independent of base model weight tensors. As explained in Section~\ref{sec:background}, expert weights in the MoE layers can be represented as tensors of shape $[M, H_{\text{out}}, H_{\text{in}}]$, where $M$ is the number of experts in the base model. One simple solution is to allocate an additional pool of adapter expert weights as tensors of shape $[M_{A}, H_{\text{out}}, H_{\text{in}}]$, where $M_{A}$ is the total number of experts in all adapters in the system. 
While attractive at first, this strategy introduces intrusive modifications to the GMM operator to load weights from both tensors, making it inferior or even infeasible in case the source code for the GMM operator is not accessible.

To keep the GMM operator intact, it is necessary to manage both base model experts and adapter experts in the same three-dimensional tensor. 
A \textit{padding} approach introduces a system-level flag $E_{\max}$ and reserves space for $E_{\max}$ experts per adapter: for a serving system that supports up to $N$ adapters in the same batch, additional space of $N\cdot E_{\max}$ experts is needed; concatenating those adapter experts with the base model along the expert dimension yields tensors of shape:
\[
[M + N \cdot E_{\max}, H_{\text{out}}, H_{\text{in}}]
\]
Note that $E_{\max}$ is a conservative configuration to be no less than the maximum number of experts for layers across all adapters.
Since most adapters fine-tune fewer than $E_{\max}$ experts per layer, this strategy can potentially lead to significant memory inefficiency.

\paragraph{Memory Fragmentation Analysis.}
We analyze real ESFT adapters to understand the memory fragmentation of the padding approach.
Due to the scarcity of off-the-shelf ESFT adapters, we choose the base model to be the ESFT vanilla model~\cite{wang2024let}, a 16B MoE model sharing the same architecture as DeepSeek-V2-Lite~\cite{deepseek_v2}, and select 10 fine-tuned ESFT adapters that cover 5 different tasks, including math~\cite{gsm8k_hf}, intent recognition~\cite{deepseek_intent_dataset}, summarization~\cite{deepseek_summary_dataset}, law~\cite{deepseek_law_dataset}, and translation~\cite{deepseek_translation_dataset}.

To quantify the deviation between the number of experts in each layer and the maximum number of experts in any layer of an adapter, we define the \textit{adapter sparsity factor} $S_i$ for adapter~$i$ as:
\[
S_{i} = \frac{\sum_{l=1}^{L} \left(E_{i} - e_i^{(l)}\right)}{L \cdot E_{i}},
\]
where $L$ is the total number of layers, $e_i^{(l)}$ is the number of experts fine-tuned in layer $l$, and $E_{i}=\max_{l=1}^{L}e_i^{(l)}$ denotes the maximum number of experts across all layers of the adapter.  
A value of $S_{i} = 0$ indicates a fully dense adapter with less inherent fragmentation, while larger values imply heavier intra-adapter padding.
Table~\ref{tab:adapter-sparsity} reports the sparsity factor for the 10 adapters: while a small number of adapters are relatively dense, most exhibit moderate to high sparsity.

Even with dense adapters, memory fragmentation can still occur as a result of inter-adapter padding, as the tensors must be padded to the largest expert count observed across all adapters.
To assess the memory efficiency of serving multiple adapters at the same time, we define the \textit{memory fragmentation factor} as:
\[
F_{\text{mem}} = \frac{L \cdot \left(M + N \cdot E_{\max}\right)}
     {\sum_{l=1}^{L} \left( M + \sum_{i=1}^{N} e_i^{(l)} \right)},
\] 
The memory fragmentation factor captures the ratio between allocated memory and actual memory usage for adapter expert weights, with $F_{\text{mem}}=1.0$ indicating no fragmentation and $F_{\text{mem}}>1.0$ indicating proportional overhead.
For the 10 ESFT adapters in Table~\ref{tab:adapter-sparsity}, the smallest feasible $E_{\max}=13$ yields an associated memory fragmentation factor of $F_{\text{mem}} = 1.51$, indicating a $51\%$ memory overhead beyond the necessary adapter weights due to padding alone.
As the memory on device is already a scarce resource, limiting this fragmentation overhead  to an acceptable degree becomes a practical challenge.

\begin{table*}[t]
\centering
\begin{tabular}{llrrr}
\toprule
\textbf{Domain} & \textbf{Adapter} & \textbf{Max.~\#Experts} & \textbf{Avg.~\#Experts} & \textbf{Sparsity} \\
\midrule
\multirow{2}{*}{Math} 
 & $\texttt{gate-math}$     & 12 & 7.04 & 0.41 \\
 & $\texttt{token-math}$    &  9 & 6.12 & 0.32 \\
\midrule
\multirow{2}{*}{Intent} 
 & $\texttt{gate-intent}$   & 12 & 9.50 & 0.21 \\
 & $\texttt{token-intent}$  &  8 & 7.12 & 0.11 \\
\midrule
\multirow{2}{*}{Summary} 
 & $\texttt{gate-summary}$  & 11 & 7.73 & 0.30 \\
 & $\texttt{token-summary}$ &  8 & 5.15 & 0.36 \\
\midrule
\multirow{2}{*}{Law} 
 & $\texttt{gate-law}$      & 12 & 7.35 & 0.39 \\
 & $\texttt{token-law}$     & 10 & 6.58 & 0.34 \\
\midrule
\multirow{2}{*}{Translation} 
 & $\texttt{gate-translation}$  & 13 & 4.69 & 0.64 \\
 & $\texttt{token-translation}$ &  6 & 3.85 & 0.36 \\
\bottomrule
\end{tabular}
\vspace{0.5em}
\caption{Expert configuration and sparsity of 10 selected ESFT adapters across 5 domains. 
\textit{Max.~\#Experts} is the maximum experts in any layer, \textit{Avg.~\#Experts} is the average experts across all layers, 
and \textit{Sparsity} is the adapter sparsity factor $S_{i}$.}
\label{tab:adapter-sparsity}
\end{table*}

\subsection{Efficient Adapter-aware Runtime}

After loading ESFT adapter experts into memory, they must be utilized accurately and efficiently at inference time, even when requests for different adapters are batched together and tokens for multiple adapters are interleaved. However, routers in the MoE layers still emit base-model expert IDs. Therefore, for a token of adapter $i$, we must transparently and deterministically redirect each selected base-model expert to its adapter-specific counterpart if available, preserving the correct semantics of ESFT. The mechanism must be lightweight as it runs on the critical path in the forward pass of LLM inference.
The mechanism must also be seamlessly integrated into existing routing and computing processes in MoE layers to preserve efficiency achieved by optimized serving engines while guaranteeing that tokens are consistently routed to the correct base-model or adapter experts, delivering task-specific outputs with minimal additional computation.

\section{\sys: Scaling Multi-ESFT Serving}
\label{sec:impl}

\subsection{Overview}
\begin{figure*}[t]
    \centering
    \includegraphics[width=\textwidth]{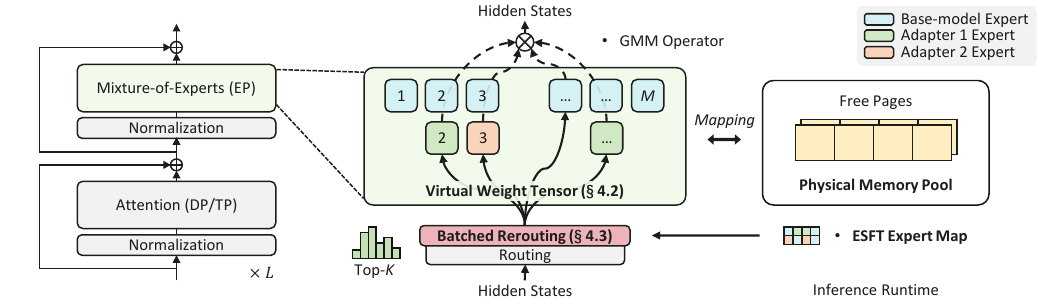}
    \caption{High-level architecture of \sys. 
    Expert weights are organized in a \textbf{virtual weight tensor} with decoupled physical page allocation managed by a \textbf{physical memory pool}. An \textbf{ESFT expert map} records the locations of fine-tuned experts within this tensor, enabling \sys to perform \textbf{batched rerouting} of tokens by replacing base-model expert IDs with their adapter-specific counterparts during inference time. 
    }
    \label{fig:overview}
\end{figure*}

We design \sys to serve multiple ESFT adapters with the following desired properties:
\begin{itemize}
\item Non-intrusive computation path modifications that enable the dynamic replacement of adapter experts while keeping the GMM operator unchanged;
\item Memory efficiency that scales with the number of active adapter experts rather than fragmentation from worst-case padding;
\item Negligible latency overhead relative to serving the base model. 
\end{itemize}

The high-level architecture of \sys is shown in Figure~\ref{fig:overview}. 
Requests are dispatched with an associated ESFT adapter ID; a request may also target the base model using a special marker. 
Tokens from requests with up to $N$ adapters can be packed together in the same batch.
Layers not fine-tuned by ESFT are kept the same as before; batched tokens participate in computations in these layers with no modifications, 
enjoying advanced optimizations employed in LLM serving systems like
continuous batching~\cite{yu2022orca,holmes2024deepspeed} and chunked prefill~\cite{agrawal2024taming,hu2024inference}.
In order to efficiently support ESFT adapters with minimal overhead and non-intrusive modifications, \sys relies on two core modules: virtual-memory-assisted expert weight management and a high-performance runtime through a batched rerouting operator.

\sys manages the weights of both base-model and adapter experts in a unified framework called \textbf{virtual weight tensor}, a single three-dimensional tensor in a contiguous virtual address space. Following the padding approach of Section~\ref{sec:challenge}, this tensor of shape $[M + N \cdot E_{\max}, H_{\text{out}}, H_{\text{in}}]$ is sized to hold the weights of the base model's $M$ experts plus the maximum possible number of experts ($E_{\max}$) for all $N$ supported adapters. While virtually contiguous, the underpinning physical memory is allocated independently and supplied on demand. 
This design provides a simple, unified view of all expert weights to the GMM operator, while avoiding much of the memory fragmentation issue mentioned in Section~\ref{sec:challenge}.

To manage expert computation at runtime, for each MoE layer $l$, \sys utilizes a \textbf{batched rerouting} operator with an \textbf{ESFT expert map}~$\Pi^{(l)}$, a two-dimensional array of shape $[N, M]$. An entry $\Pi^{(l)}[i, j]$ stores the index of expert~$j$ of adapter~$i$ in the virtual weight tensor: If expert $j$ is not fine-tuned by adapter $i$, $\Pi^{(l)}[i, j]$ simply holds the original index of $j$; if it is fine-tuned, $\Pi^{(l)}[i, j]$ points to the specific location where the adapter expert weights are loaded in the virtual weight tensor:
\[
\Pi^{(l)}[i,j] \;=\;
\begin{cases}
j, & \text{if $j$ is not fine-tuned in adapter $i$},\\
\Delta_i + \delta^{(l)}_{ij}, & \text{if $j$ is fine-tuned in adapter $i$}
\end{cases}
\]
where $\Delta_i = M + i \cdot E_{\max}$ is the offset of adapter $i$ in the first dimension of the virtual weight tensor and $\delta^{(l)}_{ij}$ is the offset of the fine-tuned expert $j$ within the range $[\Delta_i:\Delta_i + E_{\max}]$ assigned to adapter $i$ where 
$0\leq \delta^{(l)}_{ij}< e_i^{(l)}$. 

At inference time, batched rerouting directs tokens to their appropriate experts by leveraging the ESFT expert map $\Pi$ to update the top-$k$ expert IDs selected by the MoE router. For a token $x$ from adapter $A(x)$, the router first selects the original set of top-$k$ experts $\mathrm{TopK}(x)$; batched rerouting then updates this set by replacing base-model expert indexes with their corresponding (fine-tuned or not) counterparts from the ESFT expert map~$\Pi$:
\[
\mathrm{TopK}'(x) := \left\{ \Pi[A(x), j] : j \in \mathrm{TopK}(x) \right\}
\]
This updated set of top-$k$ experts, $\mathrm{TopK}'(x)$, is then passed to the unmodified inference path with token dispatching and the GMM operator, which is executed on the virtual weight tensor without needing to be aware of the adapter-specific logic.

\subsection{Virtual-Memory-Assisted Expert Weight Management}
\label{subsec:virtual_weight_memory}

\begin{figure*}[t]
    \centering
    \includegraphics[width=\textwidth]{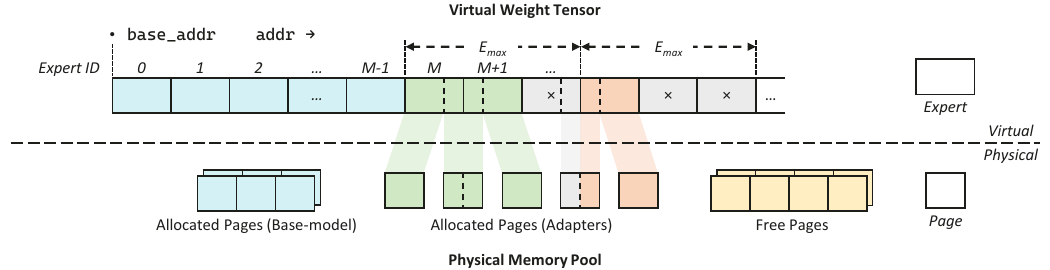}
    \caption{An illustration of a virtual weight tensor unfolded alongside its first dimension and a physical memory pool. Mappings for base-model experts (blue) are omitted. Adapter experts are padded to  $E_{\max}=3$ in the virtual address space, but only active experts (green and orange) are mapped with physical pages. Regions marked with "$\times$" (gray) represent padding without physical pages. Misalignment between expert and page boundaries leads to partially filled pages. Free pages are available for future allocations.}
    \label{fig:virtual_tensor}
\end{figure*}

As described in Section~\ref{sec:challenge}, loading adapter expert weights following the padding approach can lead to severe memory fragmentation on device. To address this issue, \sys adopts a strategy that \textit{decouples virtual memory reservation from physical memory allocation}.

Specifically, \sys reserves a contiguous virtual address space for the full weight tensor, but only allocates physical pages for the experts that are actually present in the adapters, with no allocation for the additional regions introduced by padding. When loading adapter $i$ with $e_i^{(l)}$ experts in layer $l$, physical memory is mapped only for the range:
\[
[\Delta_i:\Delta_i + e_i^{(l)}]
\]
The leftover padding range $[\Delta_i + e_i^{(l)}:\Delta_i + E_{\max}]$ is kept intentionally not mapped, ensuring that no physical memory is wasted on unused areas of the tensor.

\paragraph{Memory Pool and Mapping.}

\sys implements the virtual weight tensor through two components:
\begin{itemize}
    \item A \textbf{physical memory pool} per device that manages physical memory pages with fixed sizes (e.g., 2~MB granularity).
    The physical memory pool pre-allocates pages from the device runtime and supplies them to the virtual weight tensor at adapter loading time; evicted adapters release their pages back to the pool, which are reused for subsequent adapters or eventually reclaimed by the device runtime.
    \item An \textbf{expert memory manager} per virtual weight tensor that handles physical pages allocated to the tensor by mapping them to the desired regions used by active experts, or the corresponding unmapping when adapters are evicted.
\end{itemize}

When a virtual weight tensor is instantiated, \sys reserves the contiguous virtual address space and returns a pointer to its base address ($\texttt{base\_addr}$) without mapping any physical memory page. Upon the loading of a range of consecutive experts, either from the base model at system initialization time or from an adapter at runtime, the expert memory manager computes the starting virtual address for the range:
\[
\texttt{start\_addr} = \texttt{base\_addr} + \Delta_i \times \texttt{expert\_size}
\]
Then, it interacts with the pool by requesting the required number of physical pages and mapping them to the appropriate offsets in the virtual address space starting at $\texttt{start\_addr}$, before copying the actual expert weights into the mapped region. Unloading a range of experts works similarly in a backward manner by unmapping the physical pages and releasing them back to the physical memory pool for subsequent reuse.

Figure~\ref{fig:virtual_tensor} illustrates a virtual weight tensor in an MoE layer~$l$ and its relationship with the physical memory pool, where the system-level $E_{\max}=3$. The $M$ experts in the base model and their associated physical pages (blue) are created at system initialization time (with the mapping omitted). There are $e_0^{(l)}=2$ experts in the first (green) adapter and $e_1^{(l)}=1$ expert in the second (orange) adapter. 
Each expert consumes 1.5 pages, and there are 3 pages mapped to the region of the 2 experts in the first adapter. Regions marked with "$\times$" (gray) represent padding within the virtual weight tensor, without physical pages assigned to these regions. Free pages in the memory pool remain available for future allocations, enabling efficient reuse.

\paragraph{Expert-Page Alignment.}
A practical challenge of expert weight management is that the memory size of a model-defined expert may not be an exact multiple of the fixed granularity of physical memory pages. This misalignment can lead to conflicts where expert boundaries in virtual address space may not coincide with physical page boundaries, leading to internal fragmentation in partially filled pages.
Figure~\ref{fig:virtual_tensor} shows an example where the \texttt{start\_addr} of the second (orange) adapter is not aligned with the page size. Inconsiderate implementations would therefore lead to runtime errors or wasted memory.

To prevent such conflicts and maximize memory utilization, \sys adopts a sub-page allocation strategy, where the unused segment of a mapped but partially filled page is intentionally made available for use by a subsequently loaded neighboring adapter. We implement this strategy through rigorous tracking of the expert-page relationship and reference counting.

\begin{table}[h]
\centering
\begin{tabular}{ll}
\toprule
\textbf{API} & \textbf{Short Description} \\
\midrule
\texttt{aclrtReserveMemAddress} & Reserve virtual memory address space. \\ \midrule
\texttt{aclrtMallocPhysical}    & Create physical memory pages on the NPU. \\
\texttt{aclrtFreePhysical}    & Free physical memory pages on the NPU. \\ \midrule
\texttt{aclrtMapMem}            & Map physical memory pages to virtual addresses. \\
\texttt{aclrtUnmapMem}          & Unmap physical memory pages from virtual addresses. \\
\bottomrule
\end{tabular}
\vspace{0.5em}
\caption{Key AscendCL APIs used in virtual-memory-assisted expert weight management.}
\label{tab:memory-apis}
\end{table}

\paragraph{Memory Management APIs.}

The key Ascend Computing Language (AscendCL) APIs enabling this approach, including reserving contiguous virtual address spaces, mapping and unmapping physical pages, and managing device-level physical memory pages, are summarized in Table~\ref{tab:memory-apis}.
As shown in Section~\ref{sec:eval}, the use of virtual-memory-assisted expert weight management introduces negligible overhead in \sys.

\subsection{Batched Rerouting}

\begin{figure}
    \centering
    \includegraphics[width=0.476\textwidth]{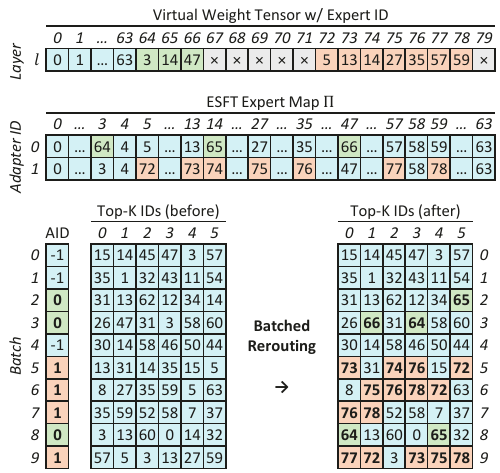}
    \caption{An example of batched rerouting with $M=64$, $K=6$, $N=2$, and $E_{\max}=8$. The original top-$k$ IDs are updated via the ESFT expert map~$\Pi$ and the adapter ID (AID) array.}
    \label{fig:batched_rerouting}
\end{figure}

To efficiently leverage the experts in the virtual weight tensor, \sys introduces a batched rerouting operator in the forward path in each MoE layer. 
After the top-$k$ expert IDs are computed by the MoE router in layer~$l$, \sys additionally reroutes some selected experts to their fine-tuned counterparts if they are for tokens in requests to different adapters, utilizing the aforementioned ESFT expert map~$\Pi^{(l)}$. This operator is performed at token granularity \cite{chen2024punica}, ensuring seamless compatibility with token-level scheduling algorithms like continuous batching~\cite{yu2022orca,holmes2024deepspeed} and chunked prefill~\cite{agrawal2024taming,hu2024inference}.

Figure~\ref{fig:batched_rerouting} shows an example of batched rerouting with a base model of $M=64$ and $K=6$, the maximum number of adapters $N=2$, and $E_{\max}=8$. In layer $l$, there are $e_0^{(l)}=3$ experts in the first adapter and $e_1^{(l)}=7$ experts in the second adapter. The hidden states of the scheduled tokens are paired with an \textit{adapter ID} (AID) array indicating the associated adapter ID of each token; a special marker value of $-1$ represents a token from requests to the base model~\cite{kwon2023vllm}. 
Each entry in the top-$k$ ID array is conditionally updated by looking up the ESFT expert map~$\Pi^{(l)}$ using both its associated AID and its value, producing an updated top-$k$ ID array corresponding to experts in the virtual weight tensor.

The batched rerouting operator is implemented as a series of vector-based operations, including broadcasting the AID array, computing offsets inside the ESFT expert map, and a gather operation. To reduce kernel launching overhead and redundant data copying, we implemented a fused kernel leveraging multiple vector cores on Ascend NPUs~\cite{liao2021ascend}.
As shown in Section~\ref{sec:eval}, the operator introduces minimal overhead for online inference and does not affect service quality.

\section{Evaluation}
\label{sec:eval}

\sys is built on top of vLLM~\cite{kwon2023vllm}, a state-of-the-art LLM inference and serving engine, and vLLM-Ascend~\cite{vllm-ascend}, a community-maintained plugin for vLLM on the Ascend~\cite{liao2021ascend} platform.

We evaluated the performance of \sys in two settings: (1) online serving with real-time dynamic workload patterns, and (2) offline batched inference for controlled microbenchmarking of system components.
We then analyze the memory efficiency of \sys and further show that \sys has zero accuracy loss in downstream tasks.

\subsection{Experimental Setup}
\label{sec:exp-setup}
\paragraph{Hardware.}
We evaluated \sys on a quad-socket server with 192 ARM cores, 1.5TB of main memory, and 8 Ascend NPUs, where each NPU has 64GB of memory.

\paragraph{Serving Framework and Setup.}
\sys is implemented on top of vLLM~\cite{kwon2023vllm} and vLLM-Ascend~\cite{vllm-ascend} with version v0.8.4. 
We use Tensor Parallelism (TP) for self-attention layers and Expert Parallelism (EP) for MoE layers when applicable.

\paragraph{Models and Adapters.}
Unlike dense models that activate the full model for each token, MoE models exhibit different expert activation patterns for different tokens.
Using synthetic adapters with dummy weights would incorrectly capture system behavior, especially under a multi-adapter setup.
We therefore match real ESFT adapters with queries from their respective domains to accurately measure the performance of \sys.
We use the 10 adapters from Table~\ref{tab:adapter-sparsity}; they are replicated for experiments beyond 10 adapters.

\paragraph{Baselines.}
To ensure fairness, we compared \sys with vLLM-Ascend on the same hardware setup. We used two baselines for vLLM-Ascend:
\begin{itemize}
 \item \textit{vLLM-Ascend (Merged)}: As there is no native support for ESFT adapters in vLLM-Ascend, we first collected merged ESFT models offline by replacing the experts in the base model with the adapter's fine-tuned experts, then served the merged model with vLLM-Ascend; for multiple ESFT adapters, we ran multiple vLLM-Ascend instances on different NPUs and dispatched requests of each domain to its instance.
 \item \textit{vLLM-Ascend (Base-Only):} When deploying multiple full merged models was infeasible due to hardware resource constraints, we instead ran a single vLLM-Ascend instance serving only the ESFT vanilla base model and sent all requests to this instance.
\end{itemize}
When unambiguous, we shorten the baseline names to \textit{vLLM-Ascend}.

\paragraph{Metrics.}
We reported prefill throughput, time-to-first-token (TTFT), decode throughput, and time-per-output-token (TPOT) as key performance metrics.

\subsection{End-to-End Performance}

\paragraph{Workloads.}
For online inference, we constructed prompts by sampling from the test sets of the datasets mentioned in Section~\ref{sec:challenge}.
Prompts from a specific dataset were sent only to adapters fine-tuned on that domain to preserve expert specialization. To evaluate \sys under workload skew, we sampled per-adapter request shares using a power-law distribution with shape parameter $\alpha$~\cite{sheng2024slora}: smaller $\alpha$ yields heavier skew (a few adapters receive most requests) while larger $\alpha$ leads to even distribution (with uniform distribution at $\alpha=1$). 
We used this distribution to assign each adapter~$i$ an arrival rate $\lambda_i$ such that the aggregate rate reached a desired value $\lambda=\sum_{i=1}^{N}\lambda_i$.
We generated one trace per adapter following a Poisson process with arrival rate $\lambda_i$, and executed all traces concurrently over a 100-second horizon.

\begin{figure*}
    \centering
    \includegraphics[width=\textwidth]{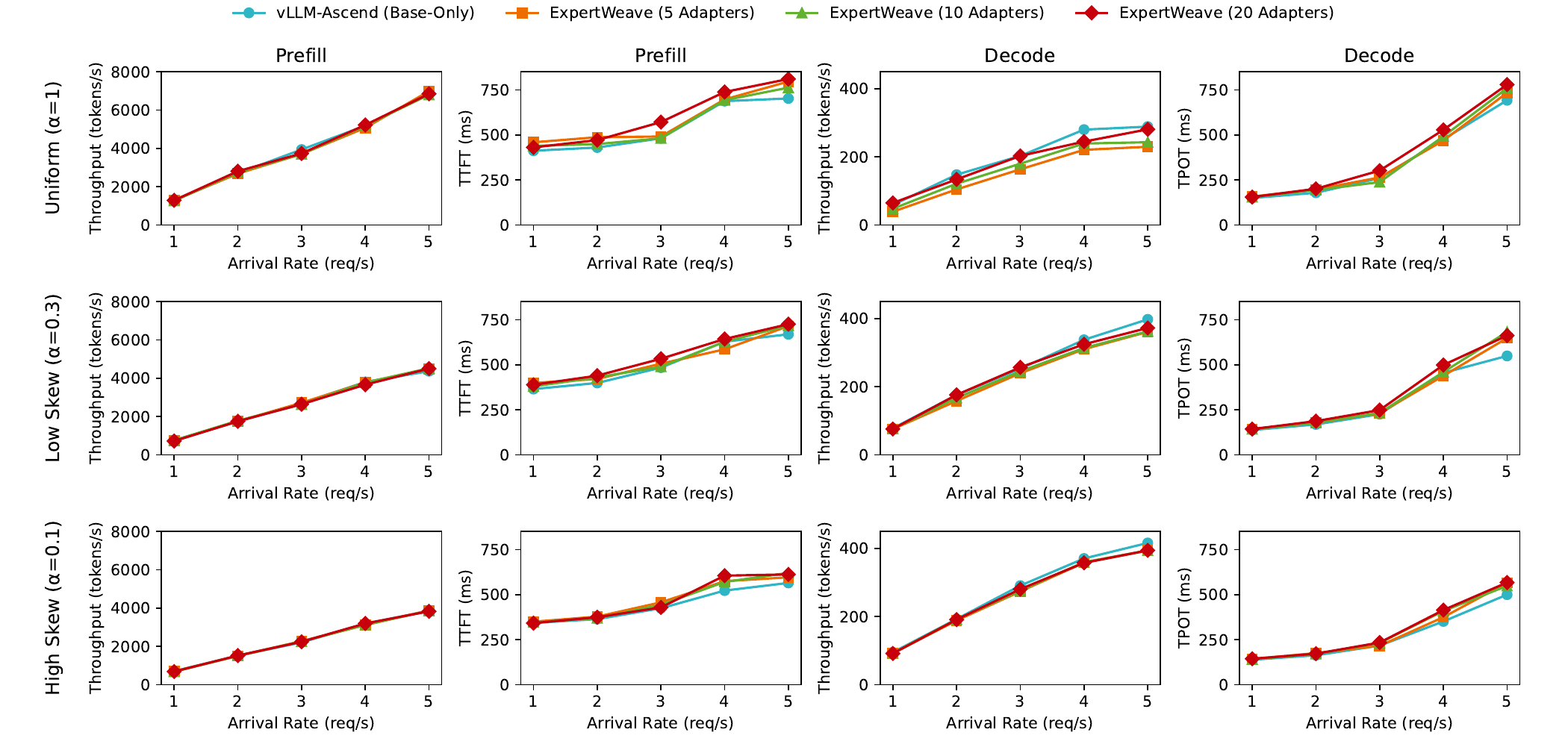}
    \caption{End-to-end performance for serving multiple adapters under uniform ($\alpha=1$) and skewed ($\alpha=0.3$ and $0.1$) workloads on 8 Ascend NPUs; \sys scales with increasing numbers of ESFT adapters with minor overhead compared with vLLM-Ascend.}
    \label{fig:e2e_performance_single_device}
\end{figure*}

\paragraph{Serving Multiple ESFT Adapters.}
To evaluate how \sys scales to the number of ESFT adapters $N$ served in the system, we performed experiments with $N=5$, 10, and 20 adapters and compared the results against the \textit{vLLM-Ascend (Base-Only)} baseline. These experiments were conducted on 8 Ascend NPUs with TP=8 (attention) and EP=8 (MoE).
The aggregate request arrival rate $\lambda$ varied from 1 to 5 requests-per-second (req/s).

As shown in Figure~\ref{fig:e2e_performance_single_device}, under a uniform workload ($\alpha=1$), serving ESFT adapters using \sys introduces only minimal TTFT overhead compared to vLLM-Ascend (Base-Only), approximately 8\% for 5 adapters and increasing modestly to about 11\% for 20 adapters. 
This overhead arises mainly from (1) batched rerouting and (2) more diverse expert activation in the GMM operator; nevertheless, this overhead remains minimal and acceptable under typical SLO requirements. 
TPOT shows a similar trend: 
Increasing the number of adapters from 5 to 10 and 20 results in an average overhead of roughly 4--11\%. 

Prefill throughput remains consistent ($<2\%$) across all configurations.
vLLM-Ascend achieves slightly higher decode throughput due to the small overhead of serving multiple ESFT adapters in \sys. 
When the number of adapters increases from 5 to 20, a slight improvement in decode throughput is observed in \sys due to the variance in output lengths in different domains.

We also observed the same qualitative behavior under skewed workloads ($\alpha=0.3$ and $0.1$); overall, \sys exhibits minimal overhead as $N$ increases, indicating effective scaling in our method.

\paragraph{Comparison with Serving Merged Models.}
Base-only deployment serves as a baseline for comparing system efficiency, but it cannot provide the same level of accuracy in downstream tasks as merged models. In this section, we therefore study how \sys compares with deploying merged models with vLLM-Ascend.
Since it is not feasible to deploy multiple merged models under reasonably comparable hardware resources and deployment strategies like TP and EP without techniques like model swapping, we instead compare the efficiency of \sys against vLLM-Ascend in a controlled setting that favors the vLLM-Ascend baseline.

We deployed \sys on two NPUs with TP=2 and EP=2. We ran two independent vLLM-Ascend instances, one for each respective merged model, and ensured the same deployment strategy by deploying each instance on two NPUs with TP=2 and EP=2 in a total of four NPUs. 
We configured \sys to use 90\% of the memory capacity via the $\texttt{gpu-memory-utilization}$ flag. 
Each vLLM-Ascend instance was restricted to 45\% of memory capacity so that aggregate memory usage was comparable while vLLM-Ascend enjoyed twice the compute resources compared to \sys.
Although this is not a fair resource setup, as we allocated more resources to vLLM-Ascend, we believe this setup is necessary as it enables us to isolate and assess the performance benefits of \sys in a multi-device deployment scenario without being affected by variance caused by deployment strategies.
In fact, we show that \sys outperforms vLLM-Ascend even under this unfair resource allocation to demonstrate its efficiency.

We used two adapters, $\texttt{gate-math}$ and $\texttt{gate-intent}$, and a fixed request arrival rate $\lambda=10$. We varied the workload by controlling the shape parameter $\alpha$.
At $\alpha = 0.32$, approximately 80\% of requests target $\texttt{gate-math}$ while the other 20\% go to $\texttt{gate-intent}$. Lowering $\alpha$ increases the skew further, with up to 95\% of requests being directed to $\texttt{gate-math}$.

\begin{figure}
    \centering
    \includegraphics[width=0.476\textwidth]{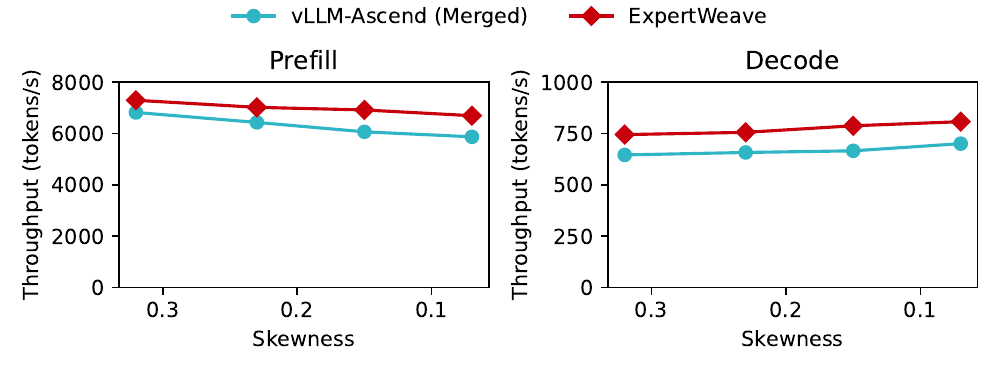}
    \caption{End-to-end performance for serving adapters vs. merged models under skewed workloads. \sys consistently outperforms vLLM-Ascend across skew levels.}
    \label{fig:e2e_performance_skewed_tp}
\end{figure}

As shown in Figure~\ref{fig:e2e_performance_skewed_tp}, \sys achieves consistently better performance with around 7--14\% higher prefill throughput and around 14--18\% higher decode throughput across various skew levels despite being allocated fewer resources compared to vLLM-Ascend. 
This performance improvement arises because vLLM-Ascend serves each merged model in isolation: as skew shifts, the NPUs used for serving $\texttt{gate-math}$ become saturated while those for $\texttt{gate-intent}$ become underutilized. This imbalance causes queuing delays, leading to increased latency and reduced throughput.
In contrast, \sys leverages resources across all available devices regardless of request distribution, sustaining high throughput.

\subsection{Microbenchmarking}
This ablation study isolates the effects of kernel fusion and virtual weight tensors on inference latency with offline microbenchmarking.

\paragraph{Workloads.}
We vary input prompt lengths to evaluate prefill latency and adjust batch sizes to evaluate decode latency. 
For prefill, we fix batch size to 1, execute queries of each prompt length 10 times, and report median TTFT.
For decode, we fix the prompt length to 1,024 tokens and decode 128 steps, and report median TPOT.
We use the $\texttt{gate-math}$ adapter and prompts from the math domain across all microbenchmarking experiments.

\begin{figure}
    \centering
    \includegraphics[width=0.476\textwidth]{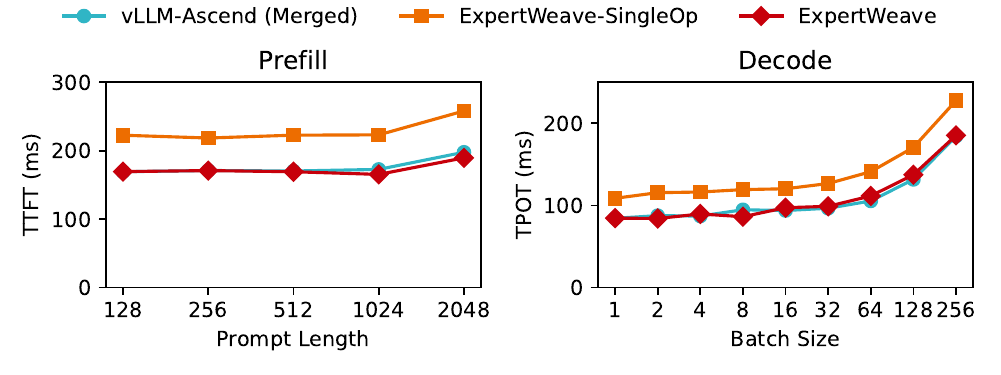}
    \caption{Performance of the batched rerouting fused kernel. \sys-SingleOp has non-negligible overhead relative to vLLM-Ascend, while \sys with fused kernel shows no observable overhead in TTFT or TPOT.}
    \label{fig:kernel_performance}
\end{figure}

\paragraph{Impact of Batched Rerouting.}
We compare the fused kernel (\textit{\sys}) 
with (1) \textit{vLLM-Ascend (Merged)} as a latency reference, and (2) \textit{\sys-SingleOp}, which implements batched rerouting using canonical PyTorch operators including \texttt{broadcast}, \texttt{gather}, etc.
Figure~\ref{fig:kernel_performance} shows that \sys-SingleOp incurred an average 29\% slowdown in TTFT and TPOT compared to vLLM-Ascend, while our fused kernel exhibited negligible ($<1\%$) overhead.

\begin{figure}
    \centering
    \includegraphics[width=0.476\textwidth]{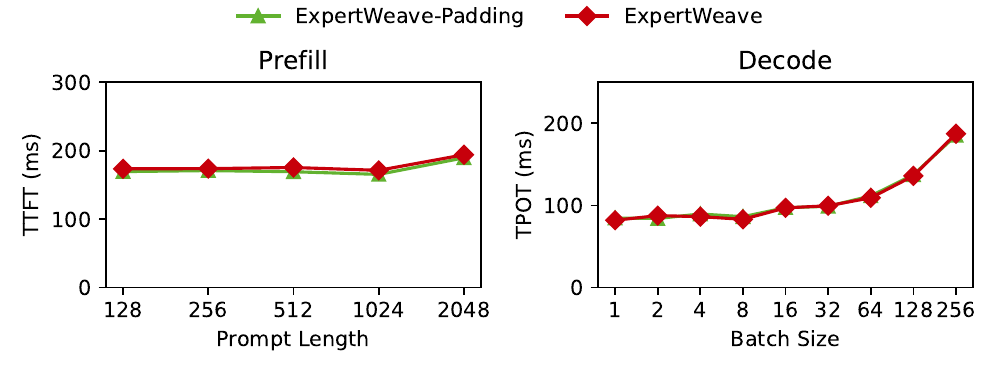}
    \caption{Effect of virtual weight tensors on TTFT and TPOT. Virtual weight tensors incur negligible overhead on inference latency.}
    \label{fig:vtensor_overhead}
\end{figure}

\paragraph{Impact of Virtual Weight Tensor.}
We evaluate the performance of virtual weight tensors (\textit{\sys}) against the padding baseline of Section~\ref{sec:challenge} (\textit{\sys-Padding}). As shown in Figure~\ref{fig:vtensor_overhead}, \sys and \sys-Padding achieve comparable TTFT ($<3\%$) and TPOT  ($<1\%$) across prompt lengths and batch sizes, demonstrating that the usage of virtual weight tensors does not degrade inference latency despite its significant memory savings and increased KV cache capacity, which we discuss in Section~\ref{sec:vtensor_memory_efficiency}.

\subsection{Memory Efficiency}
\label{sec:vtensor_memory_efficiency}
We evaluate the memory efficiency of \sys by comparing its use of virtual weight tensors with (1) \textit{vLLM-Ascend (Merged)} and (2) \textit{\sys-Padding}. We serve multiple adapters on a single NPU with 64GB of memory. For adapters, we use $\texttt{gate-math}$, $\texttt{token-math}$ and $\texttt{gate-intent}$ for our experiments.

\begin{figure}
    \centering
    \includegraphics[width=0.476\textwidth]{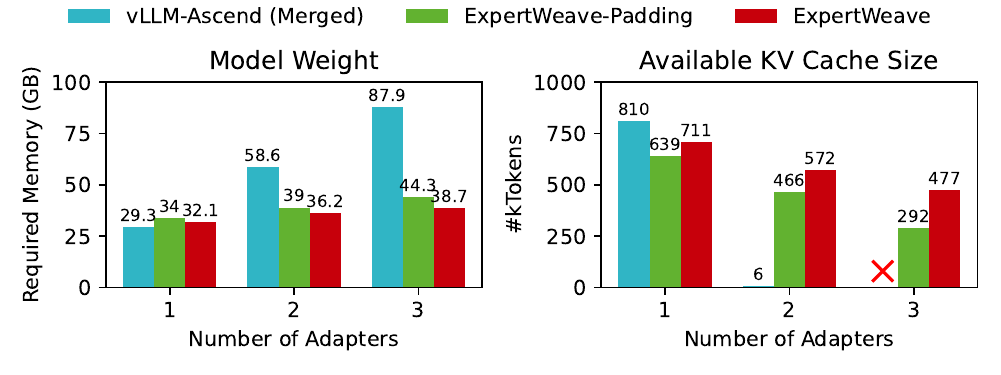}
    \caption{The memory usage of padding and the virtual weight tensor. \sys reduces the memory fragmentation, enabling larger KV cache space for concurrent requests.}
    \label{fig:memory_efficiency}
\end{figure}

As illustrated in Figure~\ref{fig:memory_efficiency}, the memory usage of vLLM-Ascend scales linearly with the number of served adapters, as it needs to deploy the full model per adapter.
While it can efficiently serve a single adapter with KV cache space for up to 810K tokens, with an additional adapter, its memory usage doubles to 58.6GB and the available space for KV cache shrinks to $\sim$6K tokens, limiting concurrency and context length. At three adapters, memory demand exceeds $\sim$88GB and the system experiences out-of-memory (OOM) errors, making this configuration infeasible.
In contrast, \sys with virtual weight tensors safely serves two adapters with a KV cache capacity of more than 572K tokens (a $94.4\times$ improvement over vLLM-Ascend) and even three adapters with 477K tokens, demonstrating its scalability.

Compared with \sys-Padding, padding alone adds 4.7GB of overhead for a single adapter, whereas \sys reduces it to 2.8GB (40.4\% reduction). The savings are 28.9\% for two adapters and 37.3\% for three adapters, enabling a 22.8--63.4\% larger KV cache than the padding-based configuration. 
These gains could lead to higher serving throughput and KV cache capacity in online serving, a critical improvement in production inference settings where memory is a limiting resource.

\subsection{Accuracy}

Finally, we show \sys has no impact on serving quality by evaluating the system on two downstream tasks: math (GSM8K~\cite{gsm8k_hf}) and intent recognition~\cite{wang2024let}. We configure \sys to load two adapters, $\texttt{gate-math}$ and $\texttt{gate-intent}$, and evaluate accuracy on both tasks. For vLLM-Ascend, we use the corresponding merged models. 

Table~\ref{tab:accuracy} shows \sys is able to serve multiple adapters while matching per-task accuracy of the respective merged model for both tasks. This result validates that \sys is a robust system for multi-adapter serving without any accuracy loss.

\begin{table}[t]
\centering
\begin{tabular}{ccc}
\toprule
                & \textbf{GSM8K}   & \textbf{Intent}   \\ \midrule
Base Model   &         56.5     &         18.6      \\ \midrule
vLLM-Ascend + $\texttt{gate-math}$   & \textbf{62.3}    & -                 \\
vLLM-Ascend + $\texttt{gate-intent}$ & -              & \textbf{78.8}     \\ \midrule
\sys            & \textbf{62.3}    & \textbf{78.8}     \\ 
\bottomrule
\end{tabular}
\vspace{0.5em}
\caption{Accuracy of \sys with two adapters and their respective merged models. \sys preserves per-task accuracy on multiple downstream tasks. The numbers are not directly comparable to those reported in \cite{wang2024let} due to differences in prompts, serving engines, hardware, etc.}
\label{tab:accuracy}
\end{table}

\section{Related Work}
\label{sec:related}

\paragraph{Advanced Multi-LoRA Serving.}
Building on top of systems like Punica~\cite{chen2024punica} and S-LoRA~\cite{sheng2024slora}, existing efforts focus on request scheduling to improve multi-LoRA serving~\cite{li2024caraserve,iliakopoulou2024chameleon,wu2024dlora}.
dLoRA~\cite{wu2024dlora} addresses skewed request distribution where a few adapters dominate request traffic by utilizing a credit-based batching algorithm to decide when to merge or unmerge adapters and a request-adapter co-migration strategy to move requests with their adapters across nodes for load balancing.
Chameleon~\cite{iliakopoulou2024chameleon} proposes an adapter-aware, non-preemptive multi-queue scheduler that avoids head-of-line (HOL) blocking.
Other works on multi-LoRA serving incorporate quantization techniques~\cite{xia2024mlgptq} and co-design with prefix caching strategies~\cite{zhang2025improving}. 
CoLD~\cite{heisler2025enhancing} combines multi-LoRA serving with contrastive decoding to enhance model performance in downstream tasks.
However, these systems are designed exclusively for LoRA adapters, which limit their adoption.

\paragraph{Serving MoE LLMs under Constrained Resources.}
Deploying large MoE models in environments with limited GPU memory is challenging due to the large memory footprint of expert weights and limited batching efficiency from dynamic routing.
MoE-Lightning~\cite{cao2025moe} proposes a CPU-GPU-I/O pipeline scheduler with paged weights to overlap data movement and computation under tight memory constraints.
MoE-Lens~\cite{yuan2025moe} introduces a resource-aware scheduler, an execution engine that overlaps prefill and decode stages and offloads attention computation to the CPU for efficient execution. 
While related, these methods are optimized for offline, throughput-oriented batch inference scenarios; our method instead targets online, latency-sensitive serving setups by sharing a single base model across multiple served adapters.

\paragraph{Virtual Memory Management for LLMs.}
Beyond the classic PagedAttention~\cite{kwon2023vllm}, recent work has explored the use of virtual memory management (VMM) APIs to improve memory management for LLMs~\cite{guo2024gmlake,prabhu2025vattention,xu2024vtensor,yu2025prism}. GMLake~\cite{guo2024gmlake} uses low-level CUDA VMM APIs to mitigate memory fragmentation during large-scale training. 
For inference workloads, vAttention~\cite{prabhu2025vattention} and vTensor~\cite{xu2024vtensor} use VMM APIs to 
manage the KV cache in single-LLM serving without relying on PagedAttention. 
Prism~\cite{yu2025prism} extends this line of work to multi-LLM serving with cross-model memory coordination to flexibly share GPU memory.
Orthogonal to virtual-memory-assisted KV cache management, our approach focuses on efficient model weight management in a multi-adapter scenario and can be combined with these approaches to further improve memory efficiency, which we leave for future work.

\section{Conclusion}
\label{sec:conclusion}

We presented \sys, a system for efficiently serving multiple Expert-Specialized Fine-Tuning (ESFT) adapters over a shared Mixture-of-Experts (MoE) base model. To improve memory utilization, \sys introduces virtual-memory-assisted expert weight management that efficiently handles the placement of base-model and adapter experts in memory, avoiding fragmentation. This design significantly reduces memory consumption compared to the vLLM-Ascend baseline that merges ESFT adapters into the base model, enabling up to $94\times$ larger KV cache capacity when serving two adapters and maintaining scalability as the number of adapters increases.
To minimize runtime overhead, \sys incorporates a fused kernel for batched rerouting, which provides lightweight adapter-aware token dispatching with negligible additional latency, while substantially outperforming unoptimized implementations.
Our evaluation further shows that \sys achieves up to $18\%$ higher throughput compared to vLLM-Ascend using comparable resources. Moreover, when serving multiple ESFT adapters concurrently, \sys preserves accuracy equivalent to the vLLM-Ascend baseline serving merged models, confirming that efficiency gains do not come at the cost of quality. Finally, scalability experiments demonstrate that \sys scales efficiently, with only a minor $4$--$11\%$ latency increase even when serving $20$ ESFT adapters.
By enabling efficient base-model sharing across ESFT adapters, \sys makes large-scale, multi-tenant deployment of specialized MoE models practical and cost-effective.

\bibliographystyle{plain}
\bibliography{paper}

\end{document}